\documentclass[aps,pra,twocolumn,reprint,superscriptaddress,floatfix]{revtex4-2}

\usepackage[T1]{fontenc}
\usepackage[utf8]{inputenc}
\usepackage{graphicx}
\usepackage{amsmath,amssymb,bm}
\usepackage{subfigure}
\usepackage{xcolor}
\usepackage[colorlinks=true,linkcolor=blue,citecolor=blue,urlcolor=blue]{hyperref}

\newcommand{\dd}{\mathrm{d}}
\newcommand{\diag}{\mathrm{diag}}

\newcommand{\Cov}{\operatorname{Cov}}

\begin{document}

\title{Quantum illumination with nonzero-mean signal–idler states via noise-enhanced heterodyne work extraction}

\author{Mustafa G\"undo\u{g}an}
\email{mustafa.guendogan@physik.hu-berlin.de}
\affiliation{Institut f\"ur Physik and Center for the Science of Materials Berlin (CSMB), Humboldt-Universit\"at zu Berlin, 12489 Berlin, Germany}

\author{Mehmet Emre Tasgin}
\email{metasgin@hacettepe.edu.tr}
\affiliation{Institute of Nuclear Sciences, Hacettepe University, 06800 Ankara, Turkey}

\date{\today}

\begin{abstract}
Room-temperature microwave and low-THz links exhibit large thermal occupations, making phase-sensitive signal–idler correlations difficult to recover after loss. We introduce a work-extraction-based quantum-illumination receiver in which the returned mode $\hat{a}_R$ is measured via heterodyne detection and the outcome is fed forward to a locally stored, possibly displaced idler. For a noisy two-mode–squeezed resource, the receiver is characterized by the heterodyne correlation parameter $x_{\rm h}=\eta c^2/[a(b+\nu_{\rm h})]$. The calibrated displaced-idler work score has Chernoff exponent $\xi_{\rm h}=x_{\rm h}/4+O(x_{\rm h}^2)$, which becomes linear in the target transmissivity $\eta$ in the weak-return, background-dominated regime, matching the leading-order performance of an \textit{ideal} OPA receiver, but achieved here via a linear and directly measurable correlation mechanism. Unlike OPA-based schemes, the present protocol does not require zero first moments and does not rely on \textit{weak-probability nonlinear} detection. In our scheme, extracted work converts hard-to-measure second-order moment correlation information into an accessible {\it first-moment signal}. Moreover, preparation noise $\bar{n}_p$, naturally present at room temperature in the microwave and THz regimes, \textit{can be directly harnessed} when correlated prior to transmission, whereas a classical coherent signal cannot utilize such incoherent thermal photons without first converting them into usable signal energy.

\end{abstract}

\maketitle

\textit{Introduction}.---Quantum illumination (QI) uses correlations between a transmitted signal and a retained idler to discriminate a weakly reflecting target from a bright noisy background. Its notable feature is that useful correlation information can remain even when the return-idler state is separable after propagation through an entanglement-breaking environment~\cite{lloyd2008enhanced,tan2008quantum,shapiro2009quantum,zhang2013entanglementBenefit,zhang2015entanglementSensing,lopaeva2013experimental,barzanjeh2015microwave,karsa2024quantum,weedbrook2016discord,nair2020fundamental,england2019quantum,gregory2020imaging,guha2011standoff,qian2023quantum,yuan2018unification}. In the microwave domain the problem is especially severe: at a few GHz and room temperature, the Planck occupation of the background can be $10^3$--$10^4$ photons per mode, so the returned marginal is dominated by thermal noise~\cite{barzanjeh2015microwave,barzanjeh2020microwave,chang2019quantum,fesquet2023perspectives,fan2018quantum}. The central experimental challenge is therefore to read out the small residual correlations without relying on highly specialized phase-conjugating or nonlinear joint receivers~\cite{guha2009gaussian,zhuang2017optimum,zhuang2017rayleigh,zhuang2017lidars,reichert2023heterohomodyne,karsa2024quantum} based on weak-probability nonlinear processes.

We develop a work-extraction readout for this setting. The receiver heterodynes the returned mode $\hat a_R$ and feeds the result forward to a local operation on the idler. The measurement information reduces the conditional uncertainty of the idler and can be converted into extractable work in the standard Gaussian measurement-and-feedback benchmark~\cite{brunelli2017detecting,manzano2018optimal,sagawa2008second,parrondo2015thermodynamics}. The scheme is written only in terms of first and second moments, so it is naturally compatible with calibrated coherent offsets and noisy Gaussian preparation.

This point is important because conventional Gaussian QI is usually formulated with zero first moments, $\langle \hat a_I\rangle=\langle \hat a_R\rangle=0$, so that the transmitter advantage is not confused with ordinary coherent illumination~\cite{tan2008quantum,shapiro2009quantum}. Here the receiver may instead use a known idler displacement as a local energetic reference. A displacement applied only to, or already present in, the retained idler does not interrogate the target and does not alter the covariance matrix, the Gaussian entanglement criterion, or the normalized return-idler covariance. Thus nonzero displaced two-mode entangled states are allowed in the protocol. The displacement is a removable baseline and, in the calibrated work channel, supplies a practical phase reference for a {\it signed linear-response observable}, see SM Sec.~5.

In contrast to optical parametric amplifier (OPA) receivers, which use a weak nonlinear interaction to access phase-sensitive correlations such as $\langle \hat a_R\hat a_I\rangle$ and are normally analyzed for zero-mean fields, the present scheme operates at the level of heterodyne data and a local idler work readout. The protocol converts the residual correlation into a deterministic, displacement-tolerant work score. This makes it a direct alternative for high-noise microwave and low-THz settings, where implementing a low-noise nonlinear joint receiver is difficult.
More importantly, work extraction provides a mechanism for mapping second order signal--idler correlations onto a signed, phase-sensitive ``first-order'' measurement outcome.

The same framework also provides a simple route to exploit trusted preparation noise. If thermal noise is present {\it before} the two-mode squeezing operation, it is amplified together with the signal–idler covariance, whereas channel noise is added only to the returned marginal. At fixed squeezing, this asymmetry enhances the normalized heterodyne correlation in a background-dominated regime. The advantage does not arise from a per-photon energy gain, but from operating in a hardware-limited setting where room-temperature thermal excitations are already present and can be {\it directly harnessed} by correlating them prior to transmission.

Preparation noise $\bar n_p$ is naturally available at room temperature in microwave and THz devices and can be used directly in this noise-enhanced scheme. A classical coherent signal cannot directly exploit those uncontrolled thermal photons as a phase-stable signal. It must first convert them into usable signal energy, which costs work. Hybrid optical--microwave generation schemes and optomechanical transduction platforms provide possible routes to preparing and probing such noisy correlated states in room-temperature or hybrid architectures~\cite{kitaeva2018generation,haase2019spontaneous,kuznetsov2020nonlinear,riedinger2018remote,abdi2015entangling,balram2017acousto,zhao2023electro}.

\textit{Gaussian model}.---We use quadratures
\begin{equation}
\hat x=(\hat a+\hat a^\dagger)/\sqrt2,\qquad
\hat p=(\hat a-\hat a^\dagger)/(i\sqrt2),
\end{equation}
so that the vacuum variance is $1/2$ and $[\hat x,\hat p]=i$. The transmitter prepares two modes by applying two-mode squeezing of strength $r$ to modes with identical trusted preparation noise
\begin{equation}
\nu_p=\bar n_p+\frac12.
\end{equation}
After the squeezing operation, the idler-signal covariance matrix is
\begin{equation}
V_{IS}=
\begin{pmatrix}
a\,\mathbb I_2 & c\,\mathbb Z\\
c\,\mathbb Z & a\,\mathbb I_2
\end{pmatrix},\qquad
\mathbb Z=\diag(1,-1),
\label{eq:VIS}
\end{equation}
with
\begin{equation}
a=\nu_p\cosh(2r),\qquad
c=\nu_p\sinh(2r).
\label{eq:ac}
\end{equation}
The signal propagates through a thermal-loss channel,
\begin{equation}
\hat a_R=\sqrt{\eta}\,\hat a_S+\sqrt{1-\eta}\,\hat a_{\rm ch},
\label{eq:channel}
\end{equation}
where $\eta$ is the transmissivity and $\nu_{\rm ch}=\bar n_{\rm ch}+1/2$ is the channel quadrature variance~\cite{Weedbrook2012}. The post-channel idler-return covariance matrix is
\begin{equation}
V_{IR}=
\begin{pmatrix}
a\,\mathbb I_2 & \sqrt{\eta}\,c\,\mathbb Z\\
\sqrt{\eta}\,c\,\mathbb Z & b\,\mathbb I_2
\end{pmatrix},
\qquad
b=\eta a+(1-\eta)\nu_{\rm ch}.
\label{eq:VIR}
\end{equation}

Possible coherent displacements are collected in the first-moment vector
\begin{equation}
\bm d_{IR}=(d_{I,x},d_{I,p},d_{R,x},d_{R,p})^T.
\end{equation}
Local displacements change $\bm d_{IR}$ but not $V_{IR}$, so they leave all covariance-based correlation and entanglement formulas unchanged~\cite{Weedbrook2012,Serafini2017Book}.

\textit{Heterodyne conditioning}.---For a Gaussian measurement on the returned mode, the conditional idler covariance is the Schur complement~\cite{giedke2002characterization,fiurasek2007gaussian,Weedbrook2012,arthurs1965simultaneous,yuen1983noise,leonhardt1997measuring}
\begin{align}
V_{I|\pi_R} &= V_I-C(V_R+\Gamma_{\pi_R})^{-1}C^T,
\label{eq:schur}
\\
\bm d_{I|\bm y_R,\pi_R}
&=\bm d_I+C(V_R+\Gamma_{\pi_R})^{-1}(\bm y_R^{\rm (h)}-\bm d_R).
\label{eq:cond-mean-general}
\end{align}
Here $\bm y_R^{({\rm h})}=(X_R^{({\rm h})},P_R^{({\rm h})})^T$ is the heterodyne outcome for the returned mode. In the present quadrature convention, ideal heterodyne detection has Gaussian seed covariance
\begin{equation}
\Gamma_{\rm h}=\nu_{\rm h}\mathbb I_2,
\qquad
\nu_{\rm h}=\frac12.
\end{equation}
Using Eq.~\eqref{eq:VIR} in Eq.~\eqref{eq:schur} gives
\begin{equation}
V_{I|{\rm h}}=
\left(a-\frac{\eta c^2}{b+\nu_{\rm h}}\right)\mathbb I_2.
\label{eq:Vcondh}
\end{equation}

\textit{Work extraction}.---In the reversible Gaussian measurement-and-feedback benchmark, the dimensionless work $w=W/k_BT$ is \textcolor{black}{the positive available work associated with a decrease of the idler nonequilibrium free energy. We use the dimensionless free-energy functional}
\begin{equation}
\textcolor{black}{f_I(\rho)\equiv \frac{F_I(\rho)}{k_BT}
=\frac{\langle \hat H_I\rangle_\rho}{k_BT}-S(\rho).}
\label{eq:free-energy}
\end{equation}
The additional work extractable due to the information gained from the measurement, corresponding to the entropy reduction term $T\, S$, contributes as
\begin{equation}
{w}_{\rm cov}
=\frac12\ln\!\left(\frac{\det V_I}{\det V_{I|{\rm h}}}\right)
=-\ln(1-x_{\rm h}),
\label{eq:Wh}
\end{equation}
where
\begin{equation}
x_{\rm h}\equiv \frac{\eta c^2}{a(b+\nu_{\rm h})},\qquad
\rho_{\rm h}\equiv \sqrt{x_{\rm h}}
=\frac{\sqrt{\eta}\,c}{\sqrt{a(b+\nu_{\rm h})}}.
\label{eq:xh}
\end{equation}
The number $\rho_{\rm h}$ is the Pearson coefficient between an idler quadrature and the corresponding heterodyne outcome after the known heterodyne noise is included. Thus $w_{\rm cov}$ is a monotone of the measured second-moment correlation. For $x_{\rm h}\ll1$, $w_{\rm cov}=x_{\rm h}+O(x_{\rm h}^2)$.

Substituting Eq.~\eqref{eq:ac} into Eq.~\eqref{eq:xh} gives
\begin{equation}
x_{\rm h}
=
\frac{\eta\,\nu_p\sinh^2(2r)}
{\cosh(2r)\left[\eta\nu_p\cosh(2r)+(1-\eta)\nu_{\rm ch}+\nu_{\rm h}\right]}.
\label{eq:xh_np}
\end{equation}
In the strongly lossy, background-dominated regime,
\begin{equation}
\eta\nu_p\cosh(2r)\ll (1-\eta)\nu_{\rm ch}+\nu_{\rm h},
\end{equation}
this reduces to
\begin{equation}
x_{\rm h}\simeq
\eta\,\frac{\nu_p}{\nu_{\rm ch}+\nu_{\rm h}}\,
\frac{\sinh^2(2r)}{\cosh(2r)}.
\label{eq:xh_weak}
\end{equation}
Thus, at fixed squeezing, the heterodyne information-related work indicator scales linearly with $\eta$ and is enhanced by the trusted preparation-noise ratio $\nu_p/(\nu_{\rm ch}+\nu_{\rm h})$. This is not a per-photon advantage at fixed transmitted brightness. It is a squeezing-limited robustness mechanism: upstream noise is correlated before the channel, whereas downstream channel noise is not. In this sense the scheme can directly use room-temperature thermal photons that are not directly usable as a classical coherent signal.

\textit{Displaced-idler work score}.---The internal-energy part of Eq.~\eqref{eq:free-energy} also contributes to the work readout. Let the harmonic idler Hamiltonian be
\begin{equation}
\hat H_I=\frac12\bm R_I^T G_H\bm R_I,
\qquad
G_H=G_H^T=\hbar\omega_I\,\mathbb I_2>0.
\label{eq:ham}
\end{equation}
In the dimensionless work $W/k_BT$, we use $G=G_H/(k_BT)$ and denote it by $G$ below. Under the correlated hypothesis $H_1$, the heterodyne-conditioned idler mean is $\bm d_I+\delta\bm d$, with
\begin{equation}
\delta\bm d_I(\bm y_R)
\equiv
\bm d_{I|\bm y_R,1}-\bm d_I
=
\frac{\sqrt{\eta}c}{b+\nu_{\rm h}}\mathbb Z \, (\bm y_R^{\rm (h)}-\bm d_R).
\label{eq:sec3-deltad}
\end{equation}
The corresponding dimensionless work benchmark is
\begin{equation}
W_1(\bm y)
=
w_{{\rm cov},1}
+
\frac12(\bm d_I+\delta\bm d)^TG(\bm d_I+\delta\bm d),
\label{eq:W1}
\end{equation}
whereas under the decorrelated hypothesis $H_0$,
\begin{equation}
W_0=
w_{{\rm cov},0}
+
\frac12\bm d_I^TG\bm d_I,
\qquad
w_{{\rm cov},0}=0.
\label{eq:W0}
\end{equation}
Thus the outcome-dependent part under $H_1$ is
\begin{equation}
\Delta W_1(\bm y)
=
\Delta w_{\rm cov}
+
\bm d_I^TG\,\delta\bm d_I
+
\frac12\delta\bm d_I^TG\delta\bm d_I.
\label{eq:DW}
\end{equation}
The quadratic term is the usual covariance-based work contribution. The term linear in $\delta\bm d_I$ appears when the idler has a nonzero first moment, either because a displaced two-mode entangled state is used or because the receiver deliberately displaces the retained idler. Its sign can be kept by a phase-stable work meter, or equivalently by subtracting the deterministic local baseline and the self-energy of the applied drive.

More importantly, the term $\bm d_I^T G\,\delta\bm d_I$ converts the
second-moment signal--idler correlations into a signed, phase-sensitive
first-order measurement observable.

\textit{Chernoff exponent}.---For each mode pair, define the standardized idler variables
\begin{equation}
Q_x=\frac{X_I-d_{I,x}}{\sqrt a},\qquad
Q_p=\frac{P_I-d_{I,p}}{\sqrt a},
\end{equation}
and the standardized heterodyne outcomes
\begin{equation}
Y_x=\frac{X_R^{({\rm h})}-d_{R,x}}{\sqrt{b+\nu_{\rm h}}},
\qquad
Y_p=\frac{P_R^{({\rm h})}-d_{R,p}}{\sqrt{b+\nu_{\rm h}}}.
\end{equation}
Under $H_1$, $\Cov(Q_x,Y_x)=\rho_{\rm h}$ and $\Cov(Q_p,Y_p)=-\rho_{\rm h}$. Under $H_0$, the same marginals are retained but the cross-covariances vanish.
Throughout this hypothesis test we use the matched-marginal, background-normalized problem, so that the receiver is benchmarked on the signal–idler correlation channel rather than on a trivial returned-energy difference.

The signed work channel $\bm d_I^TG\,\delta\bm d_I$ in Eq.~\eqref{eq:DW} generated by heterodyne feed-forward can be written as
\begin{equation}
s_m=-\lambda\left(Q_{x,m}Y_{x,m}-Q_{p,m}Y_{p,m}\right),
\label{eq:workscore}
\end{equation}
where $\lambda$ is a real gain that fixes the physical work scale and meter range. It cancels from the leading ideal exponent.

For the $x$ pair, let
\begin{equation}
\bm z=(Q_x,Y_x)^T.
\end{equation}
The Chernoff coefficient is
\begin{equation}
{\cal Q}_s
=
\int_{\mathbb R^2}\dd^2 z\,
p_0(\bm z)^{1-s}p_1(\bm z)^s,
\qquad
0\le s\le1.
\label{eq:sec3-Qs-def}
\end{equation}
For zero-mean normal distributions
\begin{equation}
p_j(\bm z)
=
\frac{1}{2\pi\sqrt{\det\Sigma_j}}
\exp\!\left(-\frac12\bm z^T\Sigma_j^{-1}\bm z\right)
\end{equation}
with
\begin{equation}
\Sigma_0=\mathbb I_2,
\qquad
\Sigma_1=
\begin{pmatrix}
1&\rho_{\rm h}\\
\rho_{\rm h}&1
\end{pmatrix},
\end{equation}
the one-quadrature-pair Chernoff exponent is
\begin{equation}
\xi_{\rm pair}(x_{\rm h})
=
-\ln\!\left[\min_{0\le s\le1}{\cal Q}_s(x_{\rm h})\right]
=
\frac{x_{\rm h}}{8}
+
\frac{7x_{\rm h}^2}{64}
+O(x_{\rm h}^3),
\label{eq:sec3-xi-pair}
\end{equation}
for $x_{\rm h}\ll1$. The $p$ pair has correlation coefficient $-\rho_{\rm h}$ and gives the same exponent because the Chernoff coefficient depends on $\rho_{\rm h}^2$. Using both heterodyne quadratures therefore gives
\begin{equation}
\xi_{\rm h}(x_{\rm h})
=
2\xi_{\rm pair}(x_{\rm h})
=
\frac{x_{\rm h}}{4}
+
\frac{7x_{\rm h}^2}{32}
+O(x_{\rm h}^3),
\label{eq:sec3-xi-h-small}
\end{equation}
which matches the leading weak-signal scaling of an ideal OPA receiver~\cite{karsa2024quantum,audenaert2007discriminating,pirandola2008computable,calsamiglia2008quantum,banchi2015quantum,wilde2017gaussian}. Combining Eqs.~\eqref{eq:xh_weak} and \eqref{eq:sec3-xi-h-small} shows explicitly that $\xi_{\rm h}\propto\eta$ in the weak-return regime.

The displaced idler is useful here because it is a local work reservoir and phase reference. It does not change the energy sent toward the target as long as only the retained idler is displaced.

\begin{figure}[t]
\centering
\subfigure[]{
\includegraphics[width=0.95\linewidth]{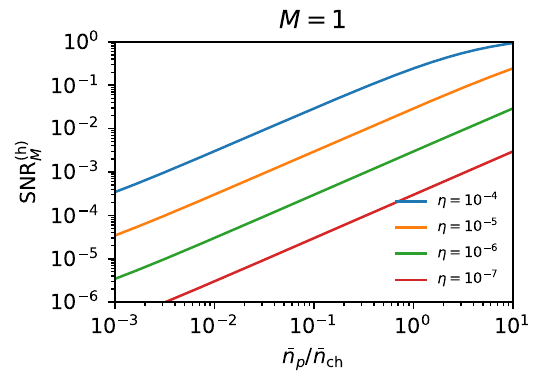}
}
\vspace{0.4em}
\subfigure[]{
\includegraphics[width=0.95\linewidth]{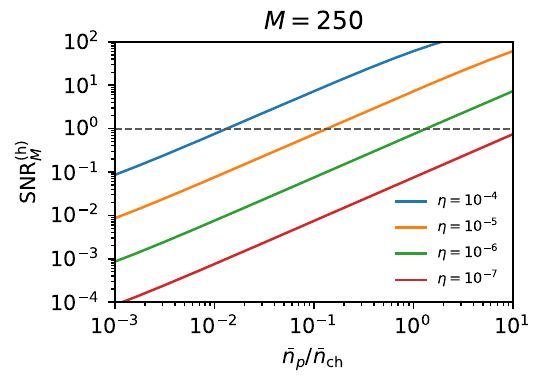}
}
\caption{Finite-sample heterodyne correlation-statistic SNR, Eq.~\eqref{eq:SNRhet}, versus the ratio of trusted preparation noise to channel background, $\bar n_p/\bar n_{\rm ch}$, for $r=4$ and $\bar n_{\rm ch}=3120$. The horizontal dashed line marks ${\rm SNR}=1$. Panel (a) uses one mode pair and panel (b) uses $M=250$ independent mode pairs. Heterodyne detection adds the fixed noise $\nu_{\rm h}=1/2$, which is negligible compared with the plotted microwave background but is retained in all formulas.}
\label{fig:snr}
\end{figure}

For the finite-sample plots below, we also use the signed heterodyne correlation statistic
\begin{equation}
C_M=\sum_{m=1}^{M}\left(Q_{x,m}Y_{x,m}-Q_{p,m}Y_{p,m}\right).
\end{equation}
It has means $\mu_1=2M\rho_{\rm h}$ and $\mu_0=0$, and variances $\sigma_1^2=2M(1+x_{\rm h})$ and $\sigma_0^2=2M$. A Monte Carlo visualization of the expected noisy-squeezing correlations underlying this statistic, including the phase scan $\rho_{\rm h}(\theta)=\rho_{\rm h}\cos(2\theta)$, is given in SM Sec.~7 and Fig.~S1. The SNR convention used in Fig.~\ref{fig:snr} is
\begin{equation}
{\rm SNR}^{({\rm h})}_M
=\frac{4(\mu_1-\mu_0)^2}{(\sigma_1+\sigma_0)^2}
=
\frac{8Mx_{\rm h}}{\left(1+\sqrt{1+x_{\rm h}}\right)^2}
\simeq 2Mx_{\rm h}.
\label{eq:SNRhet}
\end{equation}

\textit{Entanglement versus useful correlations}.---For the covariance matrix in Eq.~\eqref{eq:VIR}, the Gaussian PPT criterion gives the entanglement condition~\cite{Simon2000PPT,Duan2000}
\begin{equation}
\eta c^2>
\left(a-\frac12\right)\left(b-\frac12\right).
\label{eq:PPT}
\end{equation}
This condition can fail while $x_{\rm h}$ remains nonzero. Hence the work-extraction signal is not an entanglement witness by itself; it is an operational measure of residual second-moment correlation. This distinction is the usual QI regime, where target information can remain accessible after environmental noise has destroyed entanglement~\cite{tan2008quantum,barzanjeh2015microwave}.

\textit{Implementation and preparation noise}.---The preparation-noise enhancement is most naturally read as a fixed-squeezing effect. If the squeezing parameter $r$ is limited by hardware, increasing $\nu_p$ before the correlating operation increases both the local variance and the inter-mode covariance, while the channel noise enters only through $b$. This asymmetry is what produces Eq.~\eqref{eq:xh_weak}. The mechanism should not be presented as an energy-fair per-photon advantage over coherent illumination at fixed transmitted brightness~\cite{shapiro2009quantum}. Rather, it uses trusted thermal excitations that are \textit{already abundant} at room temperature in microwave and low-THz devices. Those excitations can be correlated by the two-mode operation and used in the receiver, whereas a classical coherent signal must first convert ambient thermal photons into phase-stable signal energy.

Two complementary implementation routes can support the noisy-state preparation considered here. First, superconducting parametric devices (e.g., Josephson mixers) enable the generation of two-mode squeezed microwave fields~\cite{Flurin2012generating,barzanjeh2015microwave,barzanjeh2020microwave}, while controlled preparation noise—introduced outside the cryogenic stage—can be injected upstream so that it is co-amplified with the signal–idler correlations. Although direct observation of microwave entanglement is often hindered by thermal fluctuations, these correlations can remain operationally accessible in quantum illumination schemes~\cite{tan2008quantum,shapiro2009quantum,barzanjeh2015microwave,karsa2024quantum}.

Second, hybrid optical–microwave/THz platforms and optomechanical transduction provide complementary routes for generating and probing correlated resources~\cite{kitaeva2018generation,haase2019spontaneous,kuznetsov2020nonlinear,riedinger2018remote,abdi2015entangling,balram2017acousto,zhao2023electro}. In such systems, correlations can be generated at optical frequencies and transferred to microwave modes—for instance, via optomechanically entangled phonon pairs converted through piezoelectric or capacitive couplings~\cite{balram2017acousto}. While thermal noise at room temperature generally obscures direct verification of microwave entanglement, useful correlations can persist and be accessed operationally, and the underlying entanglement may still be inferred indirectly from the nonclassical properties of the associated optical fields. These hybrid approaches therefore offer viable pathways toward room-temperature implementations of correlated microwave and THz resources.

\textit{Discussions \& Conclusions}.---We have introduced a work-extraction-based quantum-illumination receiver for discriminating target presence and absence in a bright thermal background. The protocol uses heterodyne detection of the returned mode and a calibrated local work readout on the idler, rather than a specialized nonlinear joint receiver. The calibrated work score has Chernoff exponent $\xi_{\rm h}=x_{\rm h}/4+O(x_{\rm h}^2)$, and since $x_{\rm h}\propto\eta$ in the background-dominated regime, the error exponent is linear in the weak return transmissivity.

Local displacement is not a problem in this receiver. A displaced two-mode entangled state, or a deliberate displacement applied only to the retained idler, leaves the target-interrogating signal energy and the covariance criteria unchanged. After calibration it supplies the local work reference needed to keep the signed correlation channel.

The use of nonzero-mean (displaced) signal–idler states enables a qualitatively different detection strategy in quantum illumination by converting otherwise inaccessible correlation information into a linear, phase-sensitive observable (\textcolor{black}{SM Sec.~5}) via measurement-induced conditional displacement of the idler. Although the displacement does not change the underlying correlations or the parameter $\textcolor{black}{x_{\rm h}}$, it effectively converts a second-order correlation structure into a \textit{first-moment signal}, producing a \textit{signed} observable proportional to $Q_xY_x-Q_pY_p$.
This reformulation yields a linear-in-$\eta$ error exponent in the weak-signal regime,  and it is naturally compatible with heterodyne detection and noise robustness. Crucially, the displacement acts as a readout resource analogous to a local oscillator, in that it converts hidden correlation information into an accessible first-order moment signal that can be directly measured and interpreted as measurement-induced work extraction. (See \textcolor{black}{SM Sec.~5} for more advantages.)

Furthermore, preparation noise $\bar{n}_p$ injected \textit{prior} to two-mode squeezing enhances the Chernoff exponent by co-amplifying inter-mode correlations, while downstream channel noise remains fixed. Although a classical signal with the same photon number can achieve similar scaling, it cannot directly exploit ambient thermal photons; instead, these must first be converted into usable energy. In contrast, the present scheme {\it directly harnesses} naturally occurring thermal noise, making it particularly suited for room-temperature microwave and low-THz platforms. Possible implementations include hybrid optical–microwave generation and optomechanical transduction architectures~\cite{kitaeva2018generation,haase2019spontaneous,kuznetsov2020nonlinear}. 
\textcolor{black}{Consequently, noise-enhanced quantum illumination is advantageous in terms of resource consumption. In implementations where the physical work-readout scale grows with the correlated idler amplitude or with the chosen feed-forward gain, the same preparation-noise mechanism can also improve the practical margin against work-meter noise; the explicit degradation factor is given in SM Sec.~3.12.}

\begin{acknowledgments}
MET gratefully thanks Ozan Ar{\i} for helpful discussions. MG acknowledges Mart\'i Perarnau-Llobet and Jens Eisert for their comments on an earlier version of this manuscript. MG is supported by Einstein Foundation Berlin through an Independent Researcher Grant and by DLR through funds provided by BMWi (OPTIMO-III, No.~50WM2347).
\end{acknowledgments}

\bibliographystyle{apsrev4-2}
\bibliography{bibliography}

@article{yuan2018unification,
	title={Unification of nonclassicality measures in interferometry},
	author={Yuan, Xiao and Zhou, Hongyi and Gu, Mile and Ma, Xiongfeng},
	journal={Physical Review A},
	volume={97},
	number={1},
	pages={012331},
	year={2018},
	publisher={APS}
}

@article{fan2018quantum,
	title={Quantum illumination using non-Gaussian states generated by photon subtraction and photon addition},
	author={Fan, Longfei and Zubairy, M Suhail},
	journal={Physical Review A},
	volume={98},
	number={1},
	pages={012319},
	year={2018},
	publisher={APS}
}

@article{qian2023quantum,
	title={Quantum induced coherence light detection and ranging},
	author={Qian, Gewei and Xu, Xingqi and Zhu, Shun-An and Xu, Chenran and Gao, Fei and Yakovlev, VV and Liu, Xu and Zhu, Shi-Yao and Wang, Da-Wei},
	journal={Physical Review Letters},
	volume={131},
	number={3},
	pages={033603},
	year={2023},
	publisher={APS}
}

@article{lloyd2008enhanced,
  title = {Enhanced Sensitivity of Photodetection via Quantum Illumination},
  author = {Lloyd, Seth},
  journal = {Science},
  volume = {321},
  number = {5895},
  pages = {1463--1465},
  year = {2008},
  doi = {10.1126/science.1160627}
}

@article{tan2008quantum,
  title = {Quantum Illumination with Gaussian States},
  author = {Tan, Si-Hui and Erkmen, Baris I. and Giovannetti, Vittorio and Guha, Saikat and Lloyd, Seth and Maccone, Lorenzo and Pirandola, Stefano and Shapiro, Jeffrey H.},
  journal = {Physical Review Letters},
  volume = {101},
  pages = {253601},
  year = {2008},
  doi = {10.1103/PhysRevLett.101.253601}
}

@article{shapiro2009quantum,
  title = {Quantum illumination versus coherent-state target detection},
  author = {Shapiro, Jeffrey H. and Lloyd, Seth},
  journal = {New Journal of Physics},
  volume = {11},
  pages = {063045},
  year = {2009},
  doi = {10.1088/1367-2630/11/6/063045}
}

@article{guha2009gaussian,
  title = {Gaussian-state quantum-illumination receivers for target detection},
  author = {Guha, Saikat and Erkmen, Baris I.},
  journal = {Physical Review A},
  volume = {80},
  pages = {052310},
  year = {2009},
  doi = {10.1103/PhysRevA.80.052310}
}

@article{zhuang2017optimum,
  title = {Optimum Mixed-State Discrimination for Noisy Entanglement-Enhanced Sensing},
  author = {Zhuang, Quntao and Zhang, Zheshen and Shapiro, Jeffrey H.},
  journal = {Physical Review Letters},
  volume = {118},
  pages = {040801},
  year = {2017},
  doi = {10.1103/PhysRevLett.118.040801}
}

@article{karsa2024quantum,
  title = {Quantum Illumination and Quantum Radar: A Brief Overview},
  author = {Karsa, Athena and Fletcher, Alasdair and Spedalieri, Gaetana and Pirandola, Stefano},
  journal = {Reports on Progress in Physics},
  volume = {87},
  pages = {094001},
  year = {2024},
  doi = {10.1088/1361-6633/ad6279}
}

@article{barzanjeh2015microwave,
  title = {Microwave Quantum Illumination},
  author = {Barzanjeh, Shabir and Guha, Saikat and Weedbrook, Christian and Vitali, David and Shapiro, Jeffrey H. and Pirandola, Stefano},
  journal = {Physical Review Letters},
  volume = {114},
  pages = {080503},
  year = {2015},
  doi = {10.1103/PhysRevLett.114.080503}
}

@article{barzanjeh2020microwave,
  title = {Microwave Quantum Illumination Using a Digital Receiver},
  author = {Barzanjeh, Shabir and Pirandola, Stefano and Vitali, David and Fink, Johannes M.},
  journal = {Science Advances},
  volume = {6},
  number = {19},
  pages = {eabb0451},
  year = {2020},
  doi = {10.1126/sciadv.abb0451}
}

@article{chang2019quantum,
  title = {Quantum-Enhanced Noise Radar},
  author = {Chang, C. W. Sandbo and Vadiraj, A. M. and Bourassa, J. and Balaji, B. and Wilson, C. M.},
  journal = {Applied Physics Letters},
  volume = {114},
  pages = {112601},
  year = {2019},
  doi = {10.1063/1.5085002}
}

@article{england2019quantum,
  title = {Quantum-Enhanced Standoff Detection Using Correlated Photon Pairs},
  author = {England, Duncan G. and Balaji, Bhashyam and Sussman, Benjamin J.},
  journal = {Physical Review A},
  volume = {99},
  pages = {023828},
  year = {2019},
  doi = {10.1103/PhysRevA.99.023828}
}

@article{gregory2020imaging,
  title = {Imaging through Noise with Quantum Illumination},
  author = {Gregory, Thomas and Moreau, Paul-Antoine and Toninelli, Ermes and Padgett, Miles J.},
  journal = {Science Advances},
  volume = {6},
  number = {6},
  pages = {eaay2652},
  year = {2020},
  doi = {10.1126/sciadv.aay2652}
}

@article{audenaert2007discriminating,
  title = {Discriminating States: The Quantum Chernoff Bound},
  author = {Audenaert, K. M. R. and Calsamiglia, J. and Mu{\~n}oz-Tapia, R. and Bagan, E. and Masanes, L. and Acin, A. and Verstraete, F.},
  journal = {Physical Review Letters},
  volume = {98},
  pages = {160501},
  year = {2007},
  doi = {10.1103/PhysRevLett.98.160501}
}

@article{pirandola2008computable,
  title = {Computable Bounds for the Discrimination of Gaussian States},
  author = {Pirandola, Stefano and Lloyd, Seth},
  journal = {Physical Review A},
  volume = {78},
  pages = {012331},
  year = {2008},
  doi = {10.1103/PhysRevA.78.012331}
}

@article{calsamiglia2008quantum,
  title = {Quantum Chernoff Bound as a Measure of Distinguishability between Density Matrices: Application to Qubit and Gaussian States},
  author = {Calsamiglia, J. and Mu{\~n}oz-Tapia, R. and Masanes, L. and Acin, A. and Bagan, E.},
  journal = {Physical Review A},
  volume = {77},
  pages = {032311},
  year = {2008},
  doi = {10.1103/PhysRevA.77.032311}
}

@article{banchi2015quantum,
  title = {Quantum Fidelity for Arbitrary Gaussian States},
  author = {Banchi, Leonardo and Braunstein, Samuel L. and Pirandola, Stefano},
  journal = {Physical Review Letters},
  volume = {115},
  pages = {260501},
  year = {2015},
  doi = {10.1103/PhysRevLett.115.260501}
}

@article{Weedbrook2012,
  title = {Gaussian Quantum Information},
  author = {Weedbrook, Christian and Pirandola, Stefano and Garc{\'i}a-Patr{\'o}n, Ra{\'u}l and Cerf, Nicolas J. and Ralph, Timothy C. and Shapiro, Jeffrey H. and Lloyd, Seth},
  journal = {Reviews of Modern Physics},
  volume = {84},
  pages = {621--669},
  year = {2012},
  doi = {10.1103/RevModPhys.84.621}
}

@book{Serafini2017Book,
  title = {Quantum Continuous Variables: A Primer of Theoretical Methods},
  author = {Serafini, Alessio},
  publisher = {CRC Press},
  address = {Boca Raton},
  year = {2017},
  doi = {10.1201/9781315118727}
}

@article{giedke2002characterization,
  title = {Characterization of Gaussian Operations and Distillation of Gaussian States},
  author = {Giedke, G{\'e}za and Cirac, J. Ignacio},
  journal = {Physical Review A},
  volume = {66},
  pages = {032316},
  year = {2002},
  doi = {10.1103/PhysRevA.66.032316}
}

@article{fiurasek2007gaussian,
  title = {Gaussian Localizable Entanglement},
  author = {Fiur{\'a}{\v{s}}ek, Jarom{\'i}r and Mi{\v{s}}ta, Ladislav},
  journal = {Physical Review A},
  volume = {75},
  pages = {060302},
  year = {2007},
  doi = {10.1103/PhysRevA.75.060302}
}

@article{arthurs1965simultaneous,
  title = {On the Simultaneous Measurement of a Pair of Conjugate Observables},
  author = {Arthurs, E. and Kelly, J. L.},
  journal = {Bell System Technical Journal},
  volume = {44},
  number = {4},
  pages = {725--729},
  year = {1965},
  doi = {10.1002/j.1538-7305.1965.tb01684.x}
}

@article{yuen1983noise,
  title = {Noise in Homodyne and Heterodyne Detection},
  author = {Yuen, H. P. and Chan, V. W. S.},
  journal = {Optics Letters},
  volume = {8},
  number = {3},
  pages = {177--179},
  year = {1983},
  doi = {10.1364/OL.8.000177}
}

@book{leonhardt1997measuring,
  title = {Measuring the Quantum State of Light},
  author = {Leonhardt, Ulf},
  publisher = {Cambridge University Press},
  address = {Cambridge},
  year = {1997}
}

@article{brunelli2017detecting,
  title = {Detecting Gaussian Entanglement via Extractable Work},
  author = {Brunelli, Matteo and Genoni, Marco G. and Barbieri, Marco and Paternostro, Mauro},
  journal = {Physical Review A},
  volume = {96},
  pages = {062311},
  year = {2017},
  doi = {10.1103/PhysRevA.96.062311}
}

@article{manzano2018optimal,
  title = {Optimal Work Extraction and Thermodynamics of Quantum Measurements and Correlations},
  author = {Manzano, Gonzalo and Plastina, Francesco and Zambrini, Roberta},
  journal = {Physical Review Letters},
  volume = {121},
  pages = {120602},
  year = {2018},
  doi = {10.1103/PhysRevLett.121.120602}
}

@article{sagawa2008second,
  title = {Second Law of Thermodynamics with Discrete Quantum Feedback Control},
  author = {Sagawa, Takahiro and Ueda, Masahito},
  journal = {Physical Review Letters},
  volume = {100},
  pages = {080403},
  year = {2008},
  doi = {10.1103/PhysRevLett.100.080403}
}

@article{parrondo2015thermodynamics,
  title = {Thermodynamics of Information},
  author = {Parrondo, Juan M. R. and Horowitz, Jordan M. and Sagawa, Takahiro},
  journal = {Nature Physics},
  volume = {11},
  pages = {131--139},
  year = {2015},
  doi = {10.1038/nphys3230}
}

@article{Simon2000PPT,
  title = {Peres-Horodecki Separability Criterion for Continuous Variable Systems},
  author = {Simon, R.},
  journal = {Physical Review Letters},
  volume = {84},
  pages = {2726--2729},
  year = {2000},
  doi = {10.1103/PhysRevLett.84.2726}
}

@article{Duan2000,
  title = {Inseparability Criterion for Continuous Variable Systems},
  author = {Duan, L.-M. and Giedke, G. and Cirac, J. I. and Zoller, P.},
  journal = {Physical Review Letters},
  volume = {84},
  pages = {2722--2725},
  year = {2000},
  doi = {10.1103/PhysRevLett.84.2722}
}

@article{fesquet2023perspectives,
  title = {Perspectives of Microwave Quantum Key Distribution in the Open Air},
  author = {Fesquet, Florian and Kronowetter, Fabian and Renger, Michael and Chen, Qiming and Honasoge, Kedar and Gargiulo, Oscar and Nojiri, Yuki and Marx, Achim and Deppe, Frank and Gross, Rudolf and Fedorov, Kirill G.},
  journal = {Physical Review A},
  volume = {108},
  pages = {032607},
  year = {2023},
  doi = {10.1103/PhysRevA.108.032607}
}

@article{Flurin2012generating,
  title = {Generating Entangled Microwave Radiation Over Two Transmission Lines},
  author = {Flurin, E. and Roch, N. and Mallet, F. and Devoret, M. H. and Huard, B.},
  journal = {Physical Review Letters},
  volume = {109},
  pages = {183901},
  year = {2012},
  doi = {10.1103/PhysRevLett.109.183901}
}

@article{kitaeva2018generation,
  title = {Generation of Optical Signal and Terahertz Idler Photons by Spontaneous Parametric Down-Conversion},
  author = {Kitaeva, G. Kh. and Kornienko, V. V. and Leontyev, A. A. and Shepelev, A. V.},
  journal = {Physical Review A},
  volume = {98},
  pages = {063844},
  year = {2018},
  doi = {10.1103/PhysRevA.98.063844}
}

@article{haase2019spontaneous,
  title = {Spontaneous Parametric Down-Conversion of Photons at 660 nm to the Terahertz and Sub-Terahertz Frequency Range},
  author = {Haase, Bj{\"o}rn and Kutas, Mirco and Riexinger, Felix and Bickert, Patricia and Keil, Andreas and Molter, Daniel and Bortz, Michael and von Freymann, Georg},
  journal = {Optics Express},
  volume = {27},
  pages = {7458--7468},
  year = {2019},
  doi = {10.1364/OE.27.007458}
}

@article{kuznetsov2020nonlinear,
  title = {Nonlinear Interference in the Strongly Nondegenerate Regime and Schmidt Mode Analysis},
  author = {Kuznetsov, Kirill A. and Malkova, Ekaterina I. and Zakharov, Roman V. and Tikhonova, Olga V. and Kitaeva, Galiya Kh.},
  journal = {Physical Review A},
  volume = {101},
  pages = {053843},
  year = {2020},
  doi = {10.1103/PhysRevA.101.053843}
}

@article{riedinger2018remote,
  title = {Remote Quantum Entanglement between Two Micromechanical Oscillators},
  author = {Riedinger, Ralf and Wallucks, Andreas and Marinkovi{\'c}, Igor and L{\"o}schnauer, Clemens and Aspelmeyer, Markus and Hong, Sungkun and Gr{\"o}blacher, Simon},
  journal = {Nature},
  volume = {556},
  pages = {473--477},
  year = {2018},
  doi = {10.1038/s41586-018-0036-z}
}

@article{abdi2015entangling,
  title = {Entangling Two Distant Non-Interacting Microwave Modes},
  author = {Abdi, Mehdi and Tombesi, Paolo and Vitali, David},
  journal = {Annalen der Physik},
  volume = {527},
  pages = {139--146},
  year = {2015},
  doi = {10.1002/andp.201400100}
}

@article{balram2017acousto,
  title = {Acousto-Optic Modulation and Optoacoustic Gating in Piezo-Optomechanical Circuits},
  author = {Balram, Krishna C. and Davan{\c c}o, Marcelo I. and Ilic, B. Robert and Kyhm, Ji-Hoon and Song, Jin Dong and Srinivasan, Kartik},
  journal = {Physical Review Applied},
  volume = {7},
  pages = {024008},
  year = {2017},
  doi = {10.1103/PhysRevApplied.7.024008}
}

@article{zhao2023electro,
  title = {Electro-Optic Transduction in Silicon via Gigahertz-Frequency Nanomechanics},
  author = {Zhao, Han and Bozkurt, Alkim and Mirhosseini, Mohammad},
  journal = {Optica},
  volume = {10},
  pages = {790--796},
  year = {2023},
  doi = {10.1364/OPTICA.479162}
}

@article{wilde2017gaussian,
  title = {Gaussian Hypothesis Testing and Quantum Illumination},
  author = {Wilde, Mark M. and Tomamichel, Marco and Lloyd, Seth and Berta, Mario},
  journal = {Physical Review Letters},
  volume = {119},
  pages = {120501},
  year = {2017},
  doi = {10.1103/PhysRevLett.119.120501}
}

@article{zhang2013entanglementBenefit,
  title = {Entanglement's Benefit Survives an Entanglement-Breaking Channel},
  author = {Zhang, Zheshen and Tengner, Maria and Zhong, Tian and Wong, Franco N. C. and Shapiro, Jeffrey H.},
  journal = {Physical Review Letters},
  volume = {111},
  pages = {010501},
  year = {2013},
  doi = {10.1103/PhysRevLett.111.010501}
}

@article{zhang2015entanglementSensing,
  title = {Entanglement-Enhanced Sensing in a Lossy and Noisy Environment},
  author = {Zhang, Zheshen and Mouradian, Sara and Wong, Franco N. C. and Shapiro, Jeffrey H.},
  journal = {Physical Review Letters},
  volume = {114},
  pages = {110506},
  year = {2015},
  doi = {10.1103/PhysRevLett.114.110506}
}

@article{lopaeva2013experimental,
  title = {Experimental Realization of Quantum Illumination},
  author = {Lopaeva, E. D. and Ruo Berchera, I. and Degiovanni, I. P. and Olivares, S. and Brida, G. and Genovese, M.},
  journal = {Physical Review Letters},
  volume = {110},
  pages = {153603},
  year = {2013},
  doi = {10.1103/PhysRevLett.110.153603}
}

@inproceedings{guha2011standoff,
  title = {Enhanced Standoff Sensing Resolution Using Quantum Illumination},
  author = {Guha, Saikat and Shapiro, Jeffrey H.},
  booktitle = {AIP Conference Proceedings},
  volume = {1363},
  pages = {113--116},
  year = {2011},
  doi = {10.1063/1.3630159}
}

@article{zhuang2017rayleigh,
  title = {Quantum Illumination for Enhanced Detection of Rayleigh-Fading Targets},
  author = {Zhuang, Quntao and Zhang, Zheshen and Shapiro, Jeffrey H.},
  journal = {Physical Review A},
  volume = {96},
  pages = {020302},
  year = {2017},
  doi = {10.1103/PhysRevA.96.020302}
}

@article{zhuang2017lidars,
  title = {Entanglement-Enhanced Lidars for Simultaneous Range and Velocity Measurements},
  author = {Zhuang, Quntao and Zhang, Zheshen and Shapiro, Jeffrey H.},
  journal = {Physical Review A},
  volume = {96},
  pages = {040304},
  year = {2017},
  doi = {10.1103/PhysRevA.96.040304}
}

@article{reichert2023heterohomodyne,
  title = {Quantum Illumination with a Hetero-Homodyne Receiver and Sequential Detection},
  author = {Reichert, Maximilian and Zhuang, Quntao and Shapiro, Jeffrey H. and Di Candia, Roberto},
  journal = {Physical Review Applied},
  volume = {20},
  pages = {014030},
  year = {2023},
  doi = {10.1103/PhysRevApplied.20.014030}
}

@article{nair2020fundamental,
  title = {Fundamental Limits of Quantum Illumination},
  author = {Nair, Ranjith and Gu, Mile},
  journal = {Optica},
  volume = {7},
  number = {7},
  pages = {771--774},
  year = {2020},
  doi = {10.1364/OPTICA.391335}
}

@article{weedbrook2016discord,
  title = {How Discord Underlies the Noise Resilience of Quantum Illumination},
  author = {Weedbrook, Christian and Pirandola, Stefano and Thompson, Jayne and Vedral, Vlatko and Gu, Mile},
  journal = {New Journal of Physics},
  volume = {18},
  pages = {043027},
  year = {2016},
  doi = {10.1088/1367-2630/18/4/043027}
}

\end{document}